\documentclass[review]{elsarticle}

\usepackage{lineno,hyperref}
\usepackage{graphicx}
\usepackage{subfig}
\usepackage{textcomp}

\modulolinenumbers[5]

\journal{Communications in Nonlinear Science and Numerical Simulation}

\begin{document}

\begin{frontmatter}

\title{Dynamics of Langton's ant allowed to periodically go straight}

\author[myaddress]{Pawe\l{} Tokarz\corref{mycorrespondingauthor}}
\cortext[mycorrespondingauthor]{Corresponding author}
\ead{paweltokarz.chem@gmail.com}

\address[myaddress]{Molecular Spectroscopy Laboratory, Faculty of Chemistry, University of Lodz, Tamka 12, 91-403 Lodz, Poland}

\begin{abstract}
A modified version of Langton's ant is considered. The modified automaton is allowed to go straight $N$-th step instead of turning. The cell state, however, is changed as usually. Depending on the value of $N$ the automaton exhibits different behaviors. Since the Cohen-Kung theorem is not applicable to this modified rule set, in most cases oscillating patterns are observed. For several values of $N$ the automaton leads to a creation of a highway. More interestingly, a few of the automata were found to exhibit a long-term chaotic behavior, exceeding even $10^{13}$ steps. The analysis of the dynamics of the system and emergent patterns is provided.
\end{abstract}

\begin{keyword}
\texttt Langton's ant\sep cellular automata \sep chaos
\end{keyword}

\end{frontmatter}


\section{Introduction}

Langton's ant is one of the most recognizable cellular automata, introduced by Langton in 1986\cite{Langton1986}. It consists of an infinite rectangular grid of cells where each is in one of two possible states -- “black” or “white”. Additionally, an agent called `an an' walks on that grid in cardinal directions (up, right, down or left). Upon entering a black cell the ant turns 90\textdegree{} clockwise and upon entering a white cell -- 90\textdegree{} counterclockwise. Then the ant switches the state of the cell and goes one step in a selected direction. Despite the set of rules is very simple the ant exhibits complex behavior. After initial symmetric and then pseudorandom movements, the ant starts to build a “highway” -- an infinite patterned strip in a diagonal direction with periodicity $\tau_h=104$ steps. Despite it has not been rigorously proven whether the ant builds the highway regardless of the finite initial configuration, it has been shown that the trajectory of Langton's ant is always unbounded -- the result known as Cohen-Kung theorem\cite{Stewart1994}.

The chaotic nature of the Langton's ant behavior found several applications, \textit{i.a.} pseudo-random number generation\cite{Hosseini2014} and image encryption\cite{Wang2015}. On the other hand, since the automaton is Turing-complete, its ability to perform calculations was used for example to solve the “Lights Out” game\cite{Arangala2016}.

Several attempts have been made to generalize the algorithm. One of the most common modifications is an introduction of more than just two states. The state of a cell is incremented every time the ant visits the cell. Whether the ant turns left or right after encountering the cell depends on a predefined rule string which binds a particular state to a particular turn\cite{Propp1994,Gale1995}.
Another modification was to introduce a special cell state -- a “grey” one -- at which the ant always goes straight and do not change that state\cite{Gale1993}. An interesting case of emergent tracks produced by ant which movement is enforced by binary sequences was reported as well\cite{Markus2006}.

\begin{figure}[h]
  \includegraphics[width=\linewidth]{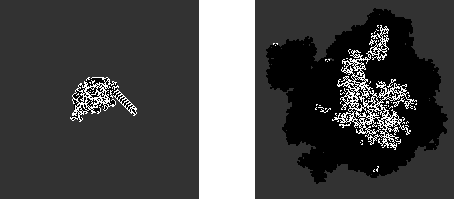}
  \caption{Left: Langton's ant after 11\,000 steps. Right: an example of proposed modified ant (6-ant) after $10^{13}$ steps. The grey color indicates the unvisited area. The ant do not distinguish between black and grey.}
  \label{fig:langton-vs-nant}
\end{figure}

Herein, a new modification is presented. The set of rules is nearly identical with the original Langton's formulation with the single exception that at every $N$\textsuperscript{th} step the ant does not turn but goes straight through the entered cell (however, the cell state is switched as usually). This simple modification appears to completely change the behavior of the ant. Since the Cohen-Kong theorem no longer holds, the movement of the ant does not have to be unbounded. In fact, several oscillating patterns have been observed. It is clear that the original Langston's ant could be obtained from this generalization by setting an infinite value of $N$. For the sake of clarity the ant characterized by a given $N$ value will be written as an $N$-ant, e.g. 6-ant.
The aim of the article is to present a general picture of the dynamics of the behavior of the $N$-ants to establish a base for a further comprehensive research (\textit{e.g.} on $N$-ant-based generators of pseud-random numbers).

\section{Calculations}

In all cases the simulation was started with a completely blank grid, with the ant oriented up. The simulations were generally run for no longer than $10^{12}$ steps with two exceptions: 6-ant was investigated up to $10^{13}$ steps and 50-ant was investigated for about $2.47\cdot10^{12}$ steps, at which point a cycle was found (see the \textit{Results} section).

The article concerns cases for $N\leq200$, leaving longer periods for further research. 

The exact algorithm is as following:

\begin{enumerate}
 \item Start with a grid filled with black cells, the ant heading up and a step counter  $c = 1$.
 \item Check the step counter. If $N\mid c$ then go directly to the point 4.
 \item If the cell state is black -- turn left; if it's white -- turn right.
 \item Switch the cell state.
 \item Move one cell forward.
 \item Increment the counter by one and go back to the point 2.
\end{enumerate}

Simulations were done on PC with Intel i3 processor. Since the algorithm for a single ant is rather not parallelizable, different $N$-ants were simulated simultaneously on different cores running other instances of the same program. The simulation itself was written in C and compiled with the GNU g++ compiler\cite{GCC}. Simulating $10^{12}$ steps took several hours, simulating $10^{13}$ steps took about 4-6 days. The density maps were generated by a script written in Processing language\cite{Processing} from data dumps generated by the C program. In all cases the sum of the visits of all cells on a dumped and read density map was checked against the actual number of steps performed by the original C program confirming the integrity of the map generation protocol.

\section{Results}


After running the simulations, it has been found that most of the tested ants eventually produce either an oscillator (meaning that after a certain number of steps $\tau_o$, different for different $N$-ants, they come back to the initial blank configuration of the board and start the cycle again) of a highway (analogically to the original Langton's ant). Several $N$-ants, however, did not produce either behavior during the steps limit ($10^{12}$). In this paper they are referred to as \textit{long-term chaotic}.

\paragraph{Highways}
The original Langton's ant (infinite $N$) can be considered as a class of $N$-ants eventually building a highway. Starting from a blank configuration the highway begins after $s_h=9977$ steps. The period of the highway is equal to $\tau_h=104$. For $N$-ants with finite $N$ the simplest highway is obviously produced by 1-ant, which never turns and immediately starts to draw a straight lane of black cells.
2-ant and 3-ant produce patterns similar to the original Langton's ant with the exception that 2-ant pattern is twice as wide and twice as high, while 3-ant pattern is just twice as high with unchanged width. Their periods are $\tau_h = 208$ and $\tau_h = 156$ respectively.

\begin{figure}[h]
\subfloat[\label{fig7:a}][$N=1$]
{\includegraphics[width=.19\linewidth]{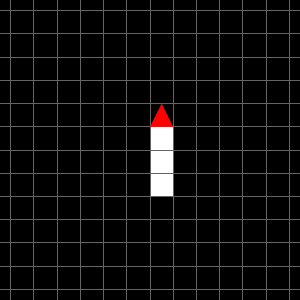}}\hfill
\subfloat[\label{fig7:b}][$N=2$]
{\includegraphics[width=.19\linewidth]{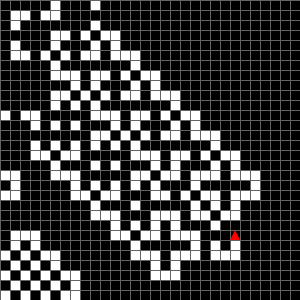}}\hfill
\subfloat[\label{fig7:b}][$N=3$]
{\includegraphics[width=.19\linewidth]{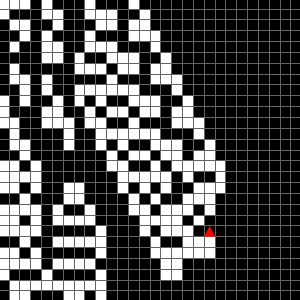}}\hfill
\subfloat[\label{fig7:b}][$N=4$]
{\includegraphics[width=.19\linewidth]{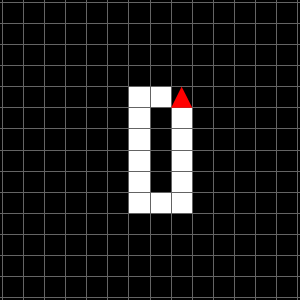}}\hfill
\subfloat[\label{fig7:b}][$N=5$]
{\includegraphics[width=.19\linewidth]{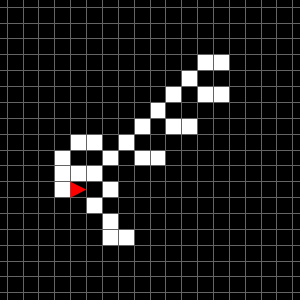}}\hfill

\subfloat[\label{fig7:b}][$N=10$]
{\includegraphics[width=.19\linewidth]{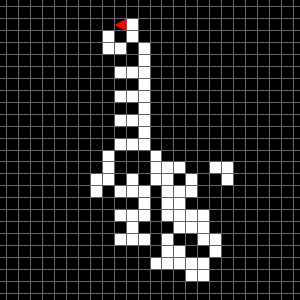}}\hfill
\subfloat[\label{fig7:b}][$N=17$]
{\includegraphics[width=.19\linewidth]{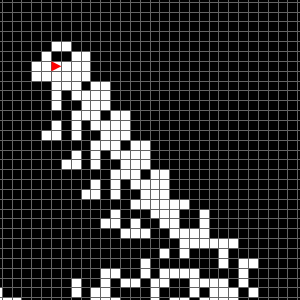}}\hfill
\subfloat[\label{fig7:b}][$N=19$]
{\includegraphics[width=.19\linewidth]{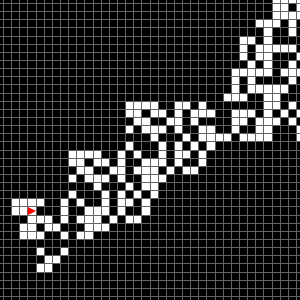}}\hfill
\subfloat[\label{fig7:b}][$N=24$]
{\includegraphics[width=.19\linewidth]{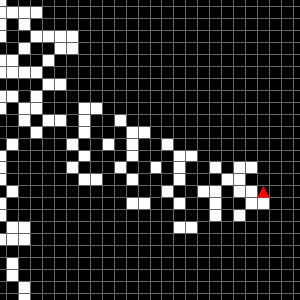}}\hfill
\subfloat[\label{fig7:d}][$N=41$]
{\includegraphics[width=.19\linewidth]{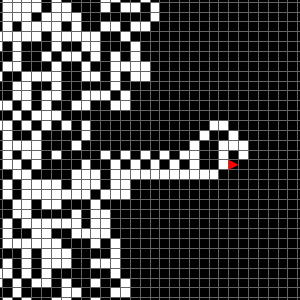}}
 \caption{All the highways found for $N<50$. The red triangle represents the ant and its direction.}
 \label{fig:highways}
 \end{figure}

The 4-ant produces a very simple, three-cells wide highway with $\tau_h=8$ that starts at fourth step and heads up.
Interestingly, the 41-ant creates a rather simple highway whose period takes only $2N$ steps ($\tau_h=82 = 2N$), but it takes as long as 280\,085\,922 steps of pseudorandom walk to enter it. Another important case is the 99-ant for which after the 100\,365\,745 steps the ant starts a highway with period equal to $N$ ($\tau_h=N=99$), which is the only such example found, excluding the trivial 1-ant). No highway was found for $N\in <100,200>$.
Graphical representations of the highways are presented on the Figure \ref{fig:highways} and the corresponding numerical data are collected in the Table \ref{table:highwayData}.

 \begin{table}[h]
 \begin{tabular}{ r r r | r r r }
  N & $s_h$ & $\tau_h$ & N & $s_h$ & $\tau_h$ \\
  \hline
  1 & 0 & 4 & 17 & 271385 & 85 \\
  2 & 20152 & 208 & 19 & 34981 & 475 \\
  3 & 15114 & 156 & 24 & 48914 & 96 \\
  4 & 4 & 8 & 41 & 280085922 & 82\\
  5 & 6 & 110 & 99 & 100365745 & 99\\
  10 & 1268 & 20 \\
 \end{tabular}
 \caption{Highways data. $s_h$ -- step at which the ant starts to build the highway; $\tau_h$ -- highway's period. More data is provided in the supplementary material.}
 \label{table:highwayData}
\end{table}


\paragraph{Oscillators}
While, regardless of the starting pattern, the Langton's ant never produces an oscillator, the evolution of most of the $N$-ants leads to oscillating patterns. This is true for at least 171 out of the first 200 $N$-ants.
Since the $N$-ant automaton is obviously reversible it is trivial to notice that any oscillating structure must periodically come back to the initial configuration. Thus, to find the periods $\tau_o$ of the oscillators it was sufficient to run each simulation until all the cells have been turned black by the ant. This was done by simply tracking the number of the white cells.
The period varies significantly -- from  $\tau_o=34\,980$ steps for $N=15$ to even $\tau_o=2\,471\,414\,288\,401$ steps in the case of $N=50$. However, even the shortest oscillator period is significantly longer than the length of the chaotic walk of the Langton's ant ($s_h=9\,977$).
No connection between $N$ and $\tau_o$ was found.

\begin{figure}[h]
  \includegraphics[width=\linewidth]{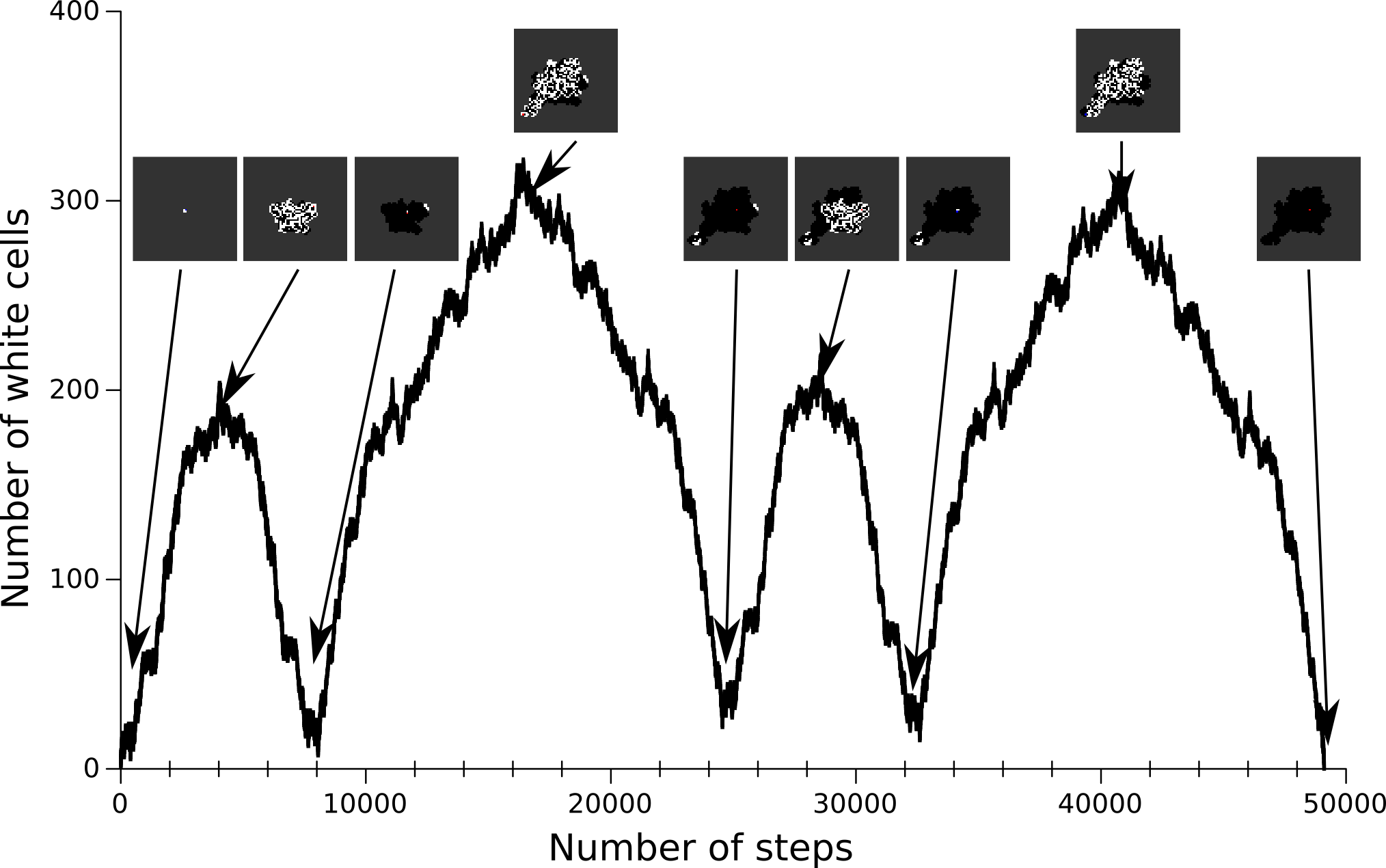}
  \caption{Evolution of a typical oscillator -- the 66-ant (see the text for explanations). More example charts are provided in a supplementary material. }
  \label{fig:oscillator}
\end{figure}

The typical evolution of the oscillating pattern consists of eight phases. The first phase is a seemingly chaotic movement, similar to that of the first 10\,000 steps of the Langton's ant. At some point $p_1$ the ant encounters a local configuration that guides the ant on the previously visited track but with the reversed direction, starting the phase 2. Due to the reversibility of the automaton, the ant erases all the white cells beside the fragment that guided it backwards, eventually at point $p_2$ leaving board almost entirely black. The ant is now at the initial position but heading a direction opposite to that it had at start. Moreover, since the ant has exactly followed backwards its own track and the automaton was started with $counter=1$ the next step after passing the initial position will correspond to the $counter=0$. Thus, the ant will start another phase with a straight move -- differently from the first phase. The third phase that has just begun is again chaotic as long as at some point $p_3$ another local configuration will guide the ant backward on its own track -- starting the fourth phase. The phase, somehow analogous to the second one, leads again to almost complete erasure of the white cells beside the first (possibly modified) and the second local guiding configuration. The phase ends at point $p_4 = \frac{1}{2}\tau_o$ with the ant at initial position and heading the same direction as at the begining of the whole evolution.
The ant starts again to behave exactly as it was in the very first phase, until it again reaches the first guiding configuration. The following four phases resemble the first four with the exception that this time the guiding configurations are not created but erased. Finally, the ant ends up on a blank array and the whole cycle starts to repeat.

\paragraph{Long-term chaotic}
During the initial systematic analysis the first fifty $N$-ants ($N\in<1,50>$) were run for at most $10^{12}$ steps. For this range of the $N$ values and within the steps limit neither period nor highway was observed for 6-ant and 50-ant. Thus, the steps limits for these two simulations were extended to $10^{13}$ steps. The 50-ant was found to form an oscillator with a period of $\tau_o=2\,471\,414\,288\,401$. Nevertheless, still neither period nor highway was observed for 6-ant. Interestingly, for these $10^{13}$ steps the 6-ant was found to reside almost exclusively inside a small 200x200 area. This result is rather exceptional for such a simple modification of the Langton's ant and suggests a possibility to exploit the automaton in further research on CA-based pesudo-random number generators. It remains unknown why exactly this ant produces such an amount of chaos, since other ants with small $N$ rather quickly oscillate or fall to a highway.
Beyond the $N=50$ the ants were simulated up to $10^{12}$ steps without exceptions. For this steps limit the long-term chaotic dynamics were observed for sixteen ants ($N=102$, 109, 128, 133, 134, 143, 150, 153, 165, 170, 184, 186, 188, 195, 197 and 198).

\paragraph{Heat maps}
To investigate the structural properties of the $N$-ants dynamics the heat maps were generated by counting the number of times the ant visited a particular cell. The data were transformed into a colored map with logarithmic scale. Blue color represents no visits, while red -- the maximum number of visits (different for each ant). For oscillating ants the map was generated after a full cycle, for ants forming the highways -- after the ant left the 600x600 board and for long-term oscillators -- after $10^{12}$ steps ($10^{13}$ in the case of 6-ant).

\begin{figure}[h]
{\includegraphics[width=.24\linewidth]{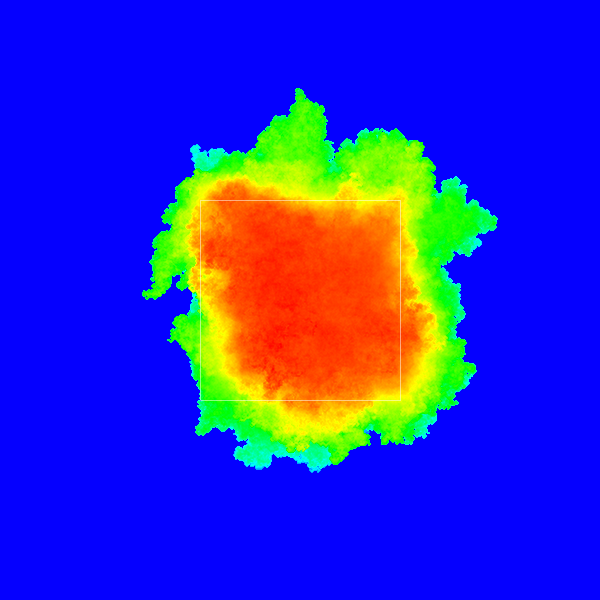}}\hfill
{\includegraphics[width=.24\linewidth]{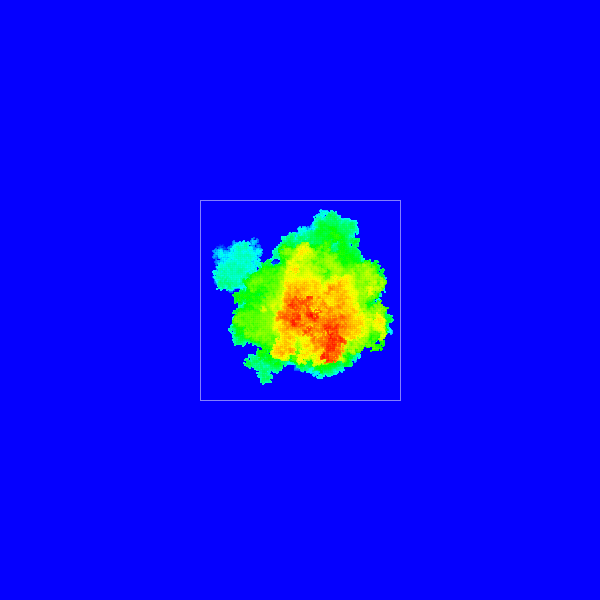}}\hfill
{\includegraphics[width=.24\linewidth]{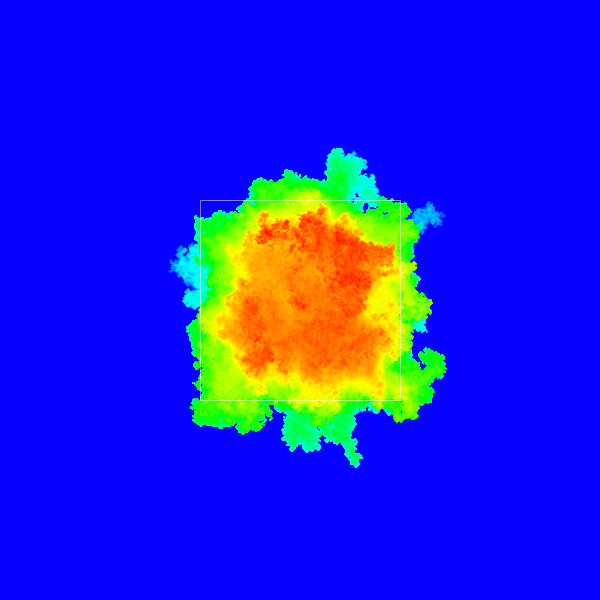}}\hfill
{\includegraphics[width=.24\linewidth]{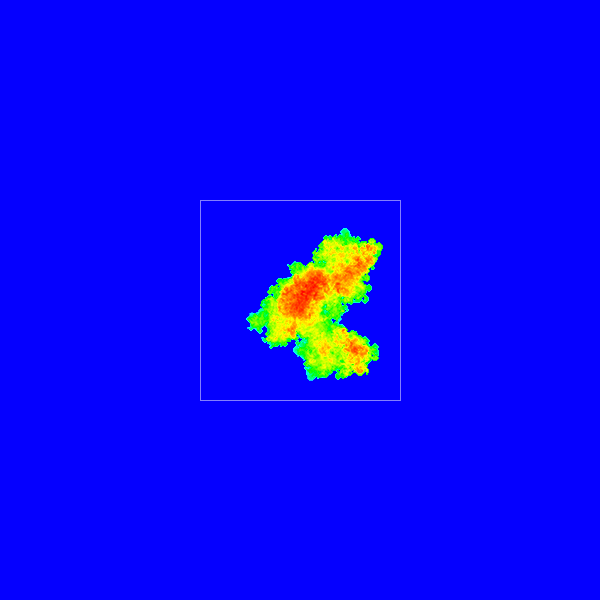}}\hfill

\subfloat[\label{fig7:a}][$N=128$ -- smooth]
{\includegraphics[width=.24\linewidth]{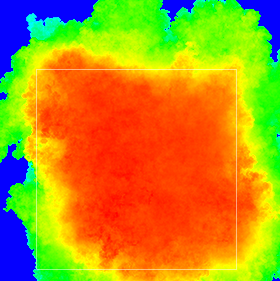}}\hfill
\subfloat[\label{fig7:b}][$N=6$ -- gridded]
{\includegraphics[width=.24\linewidth]{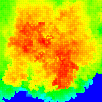}}\hfill
\subfloat[\label{fig7:b}][$N=50$ -- cloudy]
{\includegraphics[width=.24\linewidth]{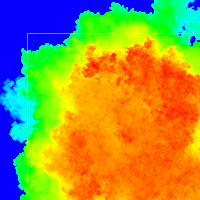}}\hfill
\subfloat[\label{fig7:b}][$N=120$ hilly]
{\includegraphics[width=.24\linewidth]{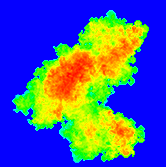}}\hfill

 \caption{Types of observed maps: smooth ($N=128$), gridded ($N=6$), cloudy ($N=50$) and hilly ($N=120$). The light rectangle marks the 200x200 area centered on the starting position.}
 \end{figure}

Several types of maps have been observed. Some ants produced smooth, fuzzy maps, suggesting that their behavior is highly isotropic, without distinguishable emergent structure. A good example is 128-ant. On the other hand there are ants (e.g. 6-ant) for which the heat maps show that the visit density follows some gridded, rectangular pattern, resembling something between a labyrinth and a chessboard. Many other ants (like 50-ant) show cloudy structures -- the map is not smooth but no grid is visible either. Finally, relatively small number of ants (like 120-ant) draws a hilly heat map with clearly visible regions with high density separated by valleys and saddles. For a given map the assignment to one of these classes might be arbitrary, since many of them exhibit mixed behaviors, however the existence of these different behaviors is clear.

Interestingly, the heat maps clearly show, that all of the $N$-ants tend to visit the same areas multiple times, even if no oscillator has been produced (yet). As an example a `coast' of the 186-ant density map is shown on a picture. Most of the border of the visited area is green, meaning that the ant has visited the border many times.

\begin{figure}[h]
  \includegraphics[width=\linewidth]{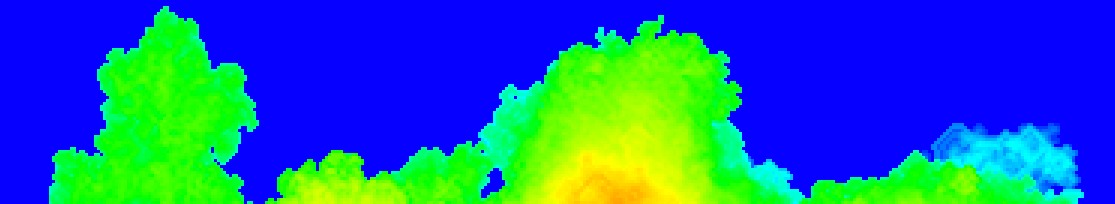}
  \caption{Part of the ‘north coast’ of the 186-ant, exhibiting typical, many times visited border and a small, rarely visited region on the right (light blue cloud).}
  \label{fig:186-coast}
\end{figure}

This green coast is adjacent to the unvisited area, meaning that there is no smooth transition between the visited and unvisited regions. On the right side of the coast there is a rarely observed example a the coast region with small number of visits, drawn in light blue.

The prevailing existence of a sharp border between the visited and unvisited regions implies that $N$-ants often follow their previous paths which makes them completely distinct from simple random walkers. This corresponds with the fact that many of them form oscillators (requiring walking backwards on own trace). On the other hand, any attempts to use $N$-ants as pseudo-random number generators this might lead to security loopholes -- even in non-oscillating regime the $N$-ants behavior might exhibit some degree of periodicity.

Despite the heat maps are useful for fast visual analysis they provide no numerical values to show exactly how often an ant walks on the border of the visited region. This might be, however, expressed as an average number of visits on the border, where the border is defined as all cells that have at least one unvisited neighbor. The data shown in Table \ref{table:borderVisits} clearly demonstrate that $N$-ants visit the border much more often than it would be expected for pure random walk. For example after $10^{12}$ steps the average number of visits at the border cells for 6-ant exceeds 36\,000 while the neighboring cells remain unvisited (Table \ref{table:borderVisits}).

 \begin{table}[h]
 \begin{tabular}{ r r r r r r }
  $N$ & $c$ & $A$ & $V_a$ & $L$ & $V_c$ \\
  \hline
  6 & $10^{12}$ & 20\,635 & 48\,461\,352.00 & 967 & 36\,557.16 \\
  7 & 6\,032\,124 & 2\,922 & 2\,064.38 & 307 & 40.25  \\
  8 & 1\,038\,238\,912 & 4\,685 & 221\,609.16 & 378 & 9\,340.88 \\
  9 & 1\,620\,252 & 1\,542 & 1\,050.75 & 193 & 47.88 \\
  50 & $10^{12}$ & 57\,216 & 17\,477\,628.00 & 1\,738 & 7\,893.48 \\
  77 & 176\,792 & 1311 & 134.85 & 174 & 9.94 \\
   &  &  &  &  &  \\
 \end{tabular}
 \caption{Example data showing the 'walk on the border' behavior. $c$ -- counter (number of steps), $A$ -- total visited area,  $L$ -- coast line length, $V_a$ -- average number of visits per cell, $V_c$ -- average number of visits per border cells.}
 \label{table:borderVisits}
\end{table}

\section{Conclusions}

Allowing the Langton's ant to periodically avoid turning at each $N$-th step significantly changes the behavior of the automaton. Among the tested $N$-ants ($N\in <1,200>$) at least 171 produce oscillating patterns with periods $\tau_o$ varying from  $\tau_o=34\,980$ ($N=15$) to $\tau_o=2\,471\,414\,288\,401$ ($N=50$). Several other $N$-ants eventually fall into a highway mode (this is especially true for small values of $N$, particularly for all cases of $N<6$). On the other hand there were found $N$-ants that generate neither oscillator nor highway for as long as $10^{12}$ steps. Nevertheless, even these long-term chaotic ants do not simply diffuse from the starting point but their movement remains confined in a small area. An exceptionally interesting case is the 6-ant, since it is the simplest long-term chaotic ant (the next one is 50-ant). It produces a pseudo-random walk for at least $10^{13}$ steps while almost exclusively residing inside a 200x200 area. Moreover, the map of the density of visits per cell for this ant exhibits a highly structured pattern.

A further research on exploiting the $N$-ants as a pseudo-random number generators could led to beneficial results due to their simplicity and relatively large amount of chaos generated, when compared to the Langton's ant. Moreover, analysis of cases of high values of $N$ might also provide interesting results, possibly especially for $N$ values exceeding the length of the chaotic walk of the Langton's ant (about $10^{4}$ steps).

\section{Acknowledgements}
The members of the 'Dzetawka' society, especially Micha\l{} Zapa\l{}a, are kindly acknowledged for fruitful discussions related to the topic. I would also like to thank to my wife, Paulina, for careful reading of the manuscript.


\bibliography{mybibfile}

\begin{thebibliography}{10}
\expandafter\ifx\csname url\endcsname\relax
  \def\url#1{\texttt{#1}}\fi
\expandafter\ifx\csname urlprefix\endcsname\relax\def\urlprefix{URL }\fi
\expandafter\ifx\csname href\endcsname\relax
  \def\href#1#2{#2} \def\path#1{#1}\fi

\bibitem{Langton1986}
C.~G. Langton, \href{http://dl.acm.org/citation.cfm?id=25201.25210}{Studying
  artificial life with cellular automata}, Phys. D 2~(1-3) (1986) 120--149.
\newline\urlprefix\url{http://dl.acm.org/citation.cfm?id=25201.25210}

\bibitem{Stewart1994}
I.~Stewart,
  \href{https://www.nature.com/scientificamerican/journal/v271/n1/pdf/scientificamerican0794-104.pdf}{The
  ultimate in anty-particles}, Mathematical Recreations 271 (1994) 104–107.
\newline\urlprefix\url{https://www.nature.com/scientificamerican/journal/v271/n1/pdf/scientificamerican0794-104.pdf}

\bibitem{Hosseini2014}
S.~M. Hosseini, H.~Karimi, M.~V. Jahan,
  \href{http://www.sciencedirect.com/science/article/pii/S2214212614000039}{Generating
  pseudo-random numbers by combining two systems with complex behaviors},
  Journal of Information Security and Applications 19~(2) (2014) 149 -- 162.
\newblock \href {http://dx.doi.org/https://doi.org/10.1016/j.jisa.2014.01.001}
  {\path{doi:https://doi.org/10.1016/j.jisa.2014.01.001}}.
\newline\urlprefix\url{http://www.sciencedirect.com/science/article/pii/S2214212614000039}

\bibitem{Wang2015}
X.~Wang, D.~Xu, \href{https://doi.org/10.1007/s11071-014-1824-0}{A novel image
  encryption scheme using chaos and langton's ant cellular automaton},
  Nonlinear Dynamics 79~(4) (2015) 2449--2456.
\newblock \href {http://dx.doi.org/10.1007/s11071-014-1824-0}
  {\path{doi:10.1007/s11071-014-1824-0}}.
\newline\urlprefix\url{https://doi.org/10.1007/s11071-014-1824-0}

\bibitem{Arangala2016}
C.~Arangala, \href{http://dx.doi.org/10.1002/cplx.21643}{Langton's turmite
  meets lights out}, Complexity 21~(5) (2016) 155--161.
\newblock \href {http://dx.doi.org/10.1002/cplx.21643}
  {\path{doi:10.1002/cplx.21643}}.
\newline\urlprefix\url{http://dx.doi.org/10.1002/cplx.21643}

\bibitem{Propp1994}
J.~Propp, \href{https://doi.org/10.1007/BF03026614}{Mathematical
  entertainments}, The Mathematical Intelligencer 16~(1) (1994) 37--44.
\newblock \href {http://dx.doi.org/10.1007/BF03026614}
  {\path{doi:10.1007/BF03026614}}.
\newline\urlprefix\url{https://doi.org/10.1007/BF03026614}

\bibitem{Gale1995}
D.~Gale, J.~Propp, S.~Sutherland, S.~Troubetzkoy,
  \href{https://doi.org/10.1007/BF03024370}{Mathematical entertainments}, The
  Mathematical Intelligencer 17~(3) (1995) 48--56.
\newblock \href {http://dx.doi.org/10.1007/BF03024370}
  {\path{doi:10.1007/BF03024370}}.
\newline\urlprefix\url{https://doi.org/10.1007/BF03024370}

\bibitem{Gale1993}
D.~Gale, \href{https://doi.org/10.1007/BF03024194}{Mathematical
  entertainments}, The Mathematical Intelligencer 15~(2) (1993) 54--58.
\newblock \href {http://dx.doi.org/10.1007/BF03024194}
  {\path{doi:10.1007/BF03024194}}.
\newline\urlprefix\url{https://doi.org/10.1007/BF03024194}

\bibitem{Markus2006}
M.~Markus, M.~Schmick, E.~Goles,
  \href{http://dx.doi.org/10.1002/cplx.20111}{Tracks emerging by forcing
  langton's ant with binary sequences}, Complexity 11~(3) (2006) 27--32.
\newblock \href {http://dx.doi.org/10.1002/cplx.20111}
  {\path{doi:10.1002/cplx.20111}}.
\newline\urlprefix\url{http://dx.doi.org/10.1002/cplx.20111}

\bibitem{GCC}
R.~M. e.~a. Stallman, \href{https://www.gnu.org/software/gcc/}{Using the gnu
  compiler collection}.
\newline\urlprefix\url{https://www.gnu.org/software/gcc/}

\bibitem{Processing}
B.~Fry, C.~Reas, \href{http://www.processing.org}{Processing library for visual
  arts and design}.
\newline\urlprefix\url{http://www.processing.org}

\end{thebibliography}

\end{document}